\documentclass[12pt]{article}
\usepackage{graphicx}
\usepackage{units}
\usepackage{subcaption}
\usepackage{chngpage}

\newcommand\pubnumber{DPF2015-21}
\newcommand\pubdate{\today}

\def\univ{Department of Physics and Astronomy\\
University of Rochester, Rochester, NY 14627 \\
{\rm and} \\
Department of Physics and Astronomy\\
Tufts University, Medford, MA 02155}

\def\Title#1{\begin{center} {\Large #1 } \end{center}}
\def\Author#1{\begin{center}{ \sc #1} \end{center}}
\def\Address#1{\begin{center}{ \it #1} \end{center}}

\newcommand\pubblock{\rightline{\begin{tabular}{l} \pubnumber\\
         \pubdate  \end{tabular}}}
\newenvironment{Abstract}{\begin{quotation}  }{\end{quotation}}
\newenvironment{Presented}{\begin{quotation} \begin{center} 
             PRESENTED AT\end{center}\bigskip 
      \begin{center}\begin{large}}{\end{large}\end{center} \end{quotation}}

\def\mnv{MINER$\nu$A}

\begin{document}
\begin{titlepage}
\pubblock

\vfill
\Title{Electron neutrino charged-current quasielastic scattering in the MINERvA experiment }
\vfill
\Author{ Jeremy Wolcott \\ for the MINERvA collaboration}
\Address{\univ}
\vfill
\begin{Abstract}
	The electron-neutrino charged-current quasielastic (CCQE) cross section on nuclei is an important input parameter to appearance-type neutrino oscillation experiments. Current experiments typically work from the muon neutrino cross section and apply corrections from theoretical arguments to obtain a prediction for the electron neutrino cross section, but to date there has been no experimental verification of the estimates for this channel at an energy scale appropriate to such experiments. We present the first measurement of an exclusive reaction in few-GeV electron neutrino interactions, namely, the cross section for a CCQE-like process, made using the MINERvA detector.  The result is given as differential cross-sections vs. the electron energy, electron angle, and square of the four-momentum transferred to the nucleus, $Q^2$.  We also compute the ratio to a muon neutrino cross-section in $Q^{2}$ from \mnv{}.  We find satisfactory agreement between this measurement and the predictions of the GENIE generator.
\end{Abstract}

\vfill

\begin{Presented}
	DPF 2015\\
	The Meeting of the American Physical Society\\
	Division of Particles and Fields\\
	Ann Arbor, Michigan, August 4--8, 2015\\
\end{Presented}

\vfill

\end{titlepage}

\def\thefootnote{\fnsymbol{footnote}}
\setcounter{footnote}{0}

	\section{Introduction}
		Current terrestrial neutrino oscillation experiments searching for fundamental information in the neutrino sector, such as the neutrino mass ordering and whether CP violation occurs for leptons, usually employ experimental designs which rely on the partial oscillation of a beam of muon neutrinos into electron neutrinos.\cite{T2K NIM,NOvA TDR}  These experiments build large detectors of heavy materials to maximize the rate of neutrino interactions, and then examine the energy distribution of the neutrinos that do interact with the detector, comparing the observed spectrum with predictions based on hypotheses of no oscillation or oscillation with given parameters.
		
		Correct prediction of the observed energy spectrum for electron neutrino interactions---on which these oscillation results depend---requires an accurate model of the rates and outgoing particle kinematics.  This, in essence, boils down to a need for precise $\nu_{e}$ cross sections on the detector materials in use.  And yet, because of the difficulties associated with producing few-GeV electron neutrino beams, even when including very recent results, only two such cross section measurements exist\cite{Gargamelle nue,T2K nue}.  Furthermore, the small statistics and inclusive nature of both of these measurements make their use as model discriminators challenging.  Instead, most simulations begin from the wealth of high-precision cross-section data available for muon neutrinos and apply corrections such as those discussed in ref. \cite{DayMcF} to obtain a prediction for $\nu_{e}$.
		
		We offer here a higher-statistics cross section for a quasielastic-like electron neutrino process, which is among the dominant reaction mechanisms at most energies of interest to oscillation experiments.  We use the \mnv{} detector, which consists of a central sampling scintillator region, built from strips of fluoror-doped scintillator glued into sheets, then stacked transverse to the beam axis; both barrel-style and downstream longitudinal electromagnetic and hadronic sampling calorimeters; and a collection of upstream passive targets of lead, iron, graphite, water, and liquid helium.  The detector design and performance are discussed in full detail elsewhere.\cite{MINERvA NIM}  \mnv{} occupies space in the NuMI $\nu_{\mu}$ beam, where it was exposed to a flux of $\sim 99$\% $\nu_{\mu}$ and $\sim 1$\% $\nu_{e}$ mostly between \unit[3-5]{GeV} for this data set.	We also compare the result for $\nu_{e}$ to a similar, previous \mnv{} result for $\nu_{\mu}$ to evaluate the assumption of the model that the only relevant difference between $\nu_{\mu}$ and $\nu_{e}$ charged-current scattering is due to the mass of the final-state charged lepton.
	
	\section{Signal definition}
		In traditional charged-current quasielastic neutrino scattering, CCQE, the neutrino is converted to a charged lepton via exchange of a W boson with a nucleon, resulting in the following reaction: $\nu_{l} n \rightarrow l^{-} p$.  (Antineutrino scattering reverses the lepton number and isospin: $\bar{\nu}_{l} p \rightarrow l^{+} n$.)  Because the \mnv{} detector is not magnetized, we cannot differentiate between electrons and positrons on an event-by-event basis.  Moreover, hadrons exiting the nucleus after the interaction can re-interact and change identity or eject other hadrons\cite{GiBUU FSI}; furthermore, pairs of nucleons correlated within the initial state may cause multiple nucleons to be ejected by a single interaction\cite{Martini corr,Nieves corr}.  Therefore, we define the signal process ``phenomenologically,'' by its final-state particles: we search for events with either an electron or positron, no other leptons or photons, any number of nucleons, and no other hadrons.  We call this type of event ``CCQE-like.''  We also demand that events originate from a 5.57-ton volume fiducial volume in the central scintillator region of \mnv{}.
		
	\section{Event selection and backgrounds}
\begin{figure}[htb]
			\centering
			
			\begin{subfigure}{0.48\textwidth}%
				\includegraphics[width=\textwidth]{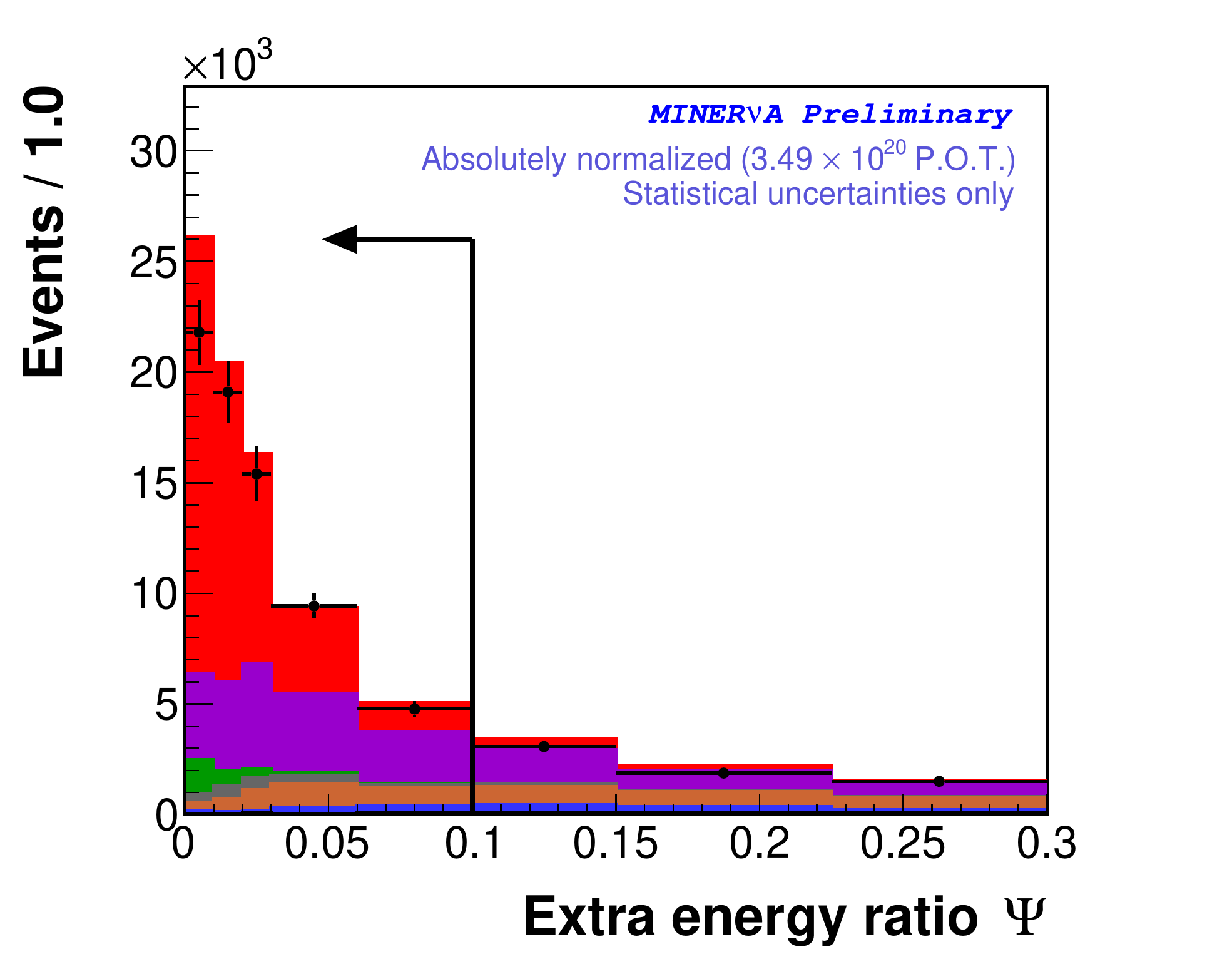}
				\caption{}
				\label{fig:psi}
			\end{subfigure}%
			\begin{subfigure}{0.48\textwidth}
				\includegraphics[width=\textwidth]{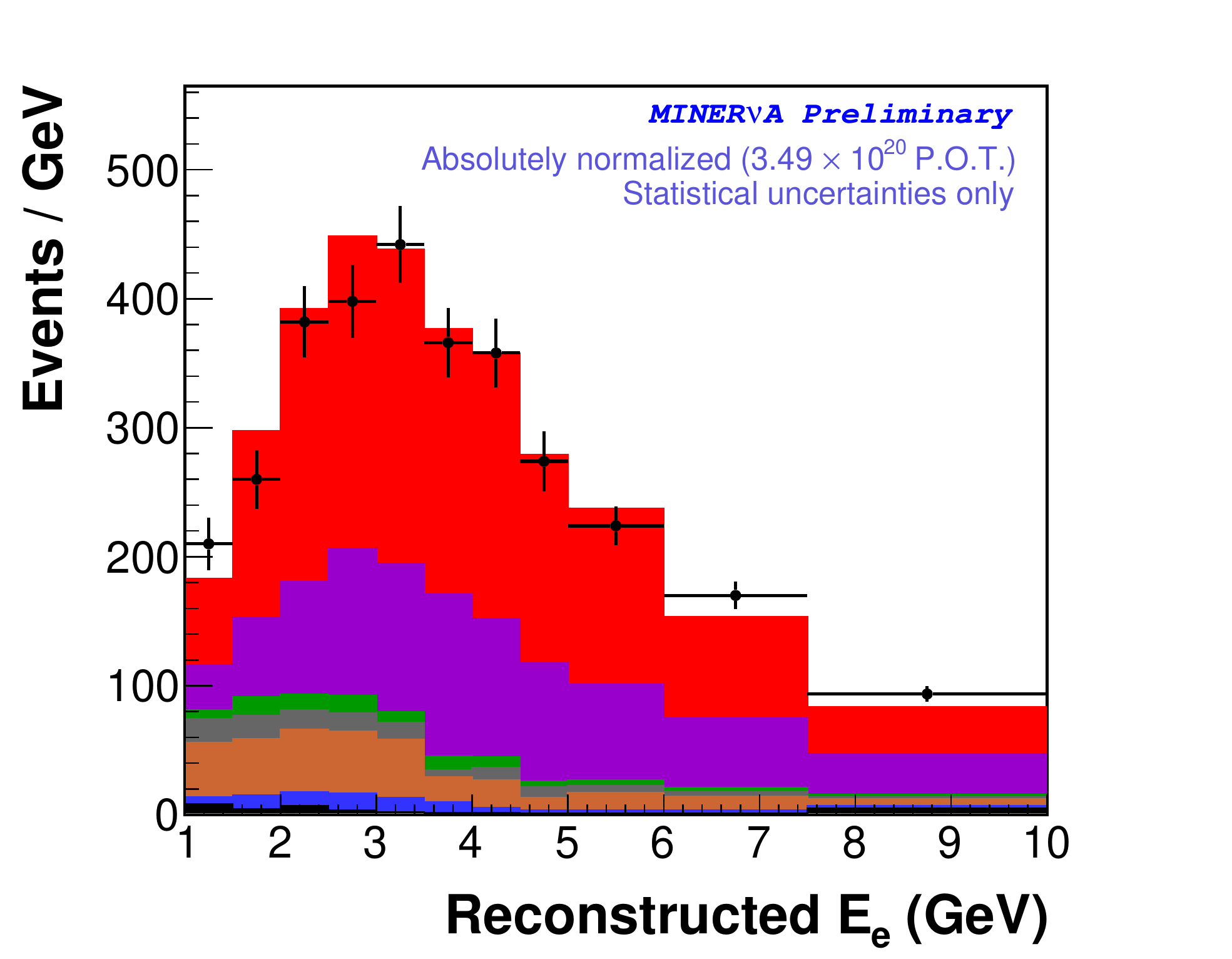}
				\caption{}
				\label{fig:selected sample}
			\end{subfigure}
			\begin{subfigure}{\textwidth}
				\centering
				\includegraphics[width=0.75\textwidth]{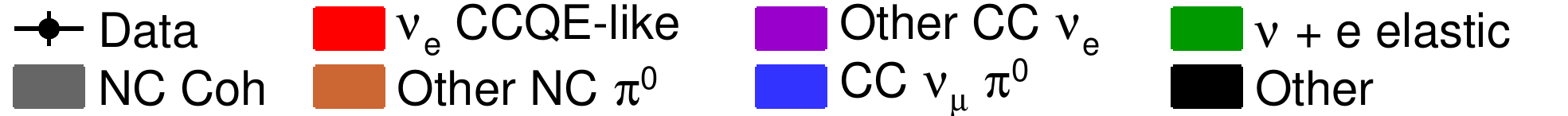}
			\end{subfigure}
			\caption{Left: example cut on $\Psi$ (defined in the text) at the most probable event visible energy, $E_{vis} = \unit[1.25]{GeV}$.  Right: event sample after all selection cuts.}
		\end{figure}

		Candidate events are selected from the data based on four major criteria.  First, a candidate must contain a reconstructed electromagnetic shower primarily contained within a cone of opening angle $7.5^{\circ}$, originating in the fiducial volume, which is identified as a shower by a multivariate PID algorithm.  The latter combines details of the energy deposition pattern both longitudinally (mean $dE/dx$, fraction of energy at downstream end of cone) and transverse to the axis of the cone (mean shower width) using a $k$-nearest-neighbors (kNN) algorithm.  Secondly, we separate electrons and positrons from photons by cutting events  in which the energy deposition rate ($dE/dx$) at the upstream end of the shower is consistent with two particles rather than one (since photons typically interact in \mnv{} by producing an electron-positron pair).  At this point, showers surviving the cuts become electron candidates.  Thirdly, we remove events with candidate muon decay electrons identified by their separation in time from the main event; these Michel electrons typically occur in inelastic interactions with final-state pions ($\pi^{\pm} \rightarrow \mu^{\pm} \rightarrow e^{\pm}$).  Our final criterion is an attempt to select CCQE-like interactions using a classifier we call ``extra energy fraction,'' $\Psi$, which, when an event's visible energy not inside the electron candidate or a sphere of radius \unit[30]{cm} centered around the cone vertex is denoted ``extra energy,'' is defined as:
		\begin{equation} \Psi = \frac{E_{\mathrm{extra}}}{E_{\mathrm{electron}}} \end{equation}
		Our cut is a function of the total visible energy of the event.  The cut at the most probable total visible energy, $E_{\mathrm{vis}} = \unit[1.25]{GeV}$, is illustrated in fig. \ref{fig:psi}.  Finally, we retain only events with reconstructed electron energy $E_{e} \geq \unit[0.5]{GeV}$ and reconstructed neutrino energy $E_{\nu}^{QE} \leq \unit[10]{GeV}$.  Here the lower bound excludes a region where the expected flux of electron-flavor neutrinos is small and the backgrounds are large, and the upper bound restricts the sample to events where the uncertainties on flux prediction are tolerable.  The distribution of events selected by this sequence is shown in fig. \ref{fig:selected sample}.
		
		As fig. \ref{fig:selected sample} shows, even after the final selection, a significant fraction of the sample is predicted to be from background processes.  To validate the background predictions from the generator, we use an \textit{in situ} \mnv{} measurement based on elastic scattering of neutrinos from atomic electrons\cite{Jaewon thesis} and a recent \mnv{} measurement of charged-current coherent pion production\cite{MINERvA coherent} to constrain the $\nu-e$ and NC coherent backgrounds.  We then attempt to constrain the remaining components of the background model by examining sidebands in two of the variables already mentioned.  The first of these is composed of events that contain Michel electron candidates, which results in a nearly pure sideband of inelastic $\nu_{e}$ events.  The second sideband is in the extra energy fraction $\Psi$; a sample of events at larger $\Psi$ constitutes a sideband rich in both the $\nu_{e}$ inelastic background and backgrounds where photon(s) from a $\pi^{0}$ decay comprise the electromagnetic shower.  We use these sidebands together to fit the normalizations of the three major backgrounds: $\nu_{e}$ inelastic events, neutral-current incoherent $\pi^{0}$ events, and charged-current incoherent $\pi^{0}$ events.  The normalizations of the $\nu_{e}$ background and the sum of the $\pi^{0}$ backgrounds are each fitted using distributions in both reconstructed candidate electron angle and energy, across the two sidebands, to obtain scale factors that represent the best estimate of the normalizations in the data as compared to the prediction from GENIE.  We obtain scale factors of $0.89 \pm 0.08$ and $1.06 \pm 0.12$, respectively.  Subsequent to the constraint, we scale the backgrounds in the signal region and subtract them from the data.  We then compare the simulated prediction of the signal process to the background-subtracted data.

	\section{Cross section result}
		We calculate three differential cross sections in electron angle, electron energy, and four-momentum transferred from neutrino to nucleus $Q^{2}$.  For $Q^{2}$, we employ the commonly-used CCQE approximations (assuming a stationary target nucleon) which allow us to compute the neutrino kinematics from just the lepton variables:
		\begin{equation}
			E_{\nu}^{QE} = \frac{m_{n}^{2} - (m_{p} - E_{b})^{2} - m_{e}^{2} + 2(m_{p}-E_{b}E_{e})}{2(m_{p} - E_{b} - E_{e} + p_{e} \cos{\theta_{e}})}
		\end{equation}
		\begin{equation}
			Q^{2}_{QE} = 2 E_{\nu}^{QE} \left(E_{e} - p_{e} \cos{\theta_{e}}\right) - m_{e}^{2}
			\label{eq:q2}
		\end{equation}
		The cross sections are calculated in bins $i$ according to the following rule for sample variable $\xi$, with $\epsilon$ representing signal acceptance, $\Phi$ the flux integrated over the energy range of the measurement, $T_{n}$ the number of targets (CH molecules) in the fiducial region, $\Delta_{i}$ the width of bin $i$, and $U_{ij}$ a matrix correcting for detector smearing in the variable of interest:
		\begin{equation}
			\label{eq:dsigma}
			\left( \frac{d\sigma}{d\xi} \right)_{i} = \frac{1}{\epsilon_{i} \Phi T_{n} \left(\Delta_{i}\right)} \times \sum_{j}{U_{ij} \left(N_{j}^{\mathrm{data}} - N_{j}^{\mathrm{bknd\ pred}}\right)}
		\end{equation}
		
		We perform unfolding in these variables using a Bayesian technique\cite{D'Agostini unf} with a single iteration.  The unfolding matrices $U_{ij}$ needed as input are predicted by our simulation.  Our prediction for the neutrino flux $\Phi$ by which we then divide is derived from a GEANT4-based simulation of the NuMI beamline (described further in ref. \cite{antinumu PRL}).  In addition, the neutrino-electron elastic scattering measurement mentioned above provides an \textit{in situ}, data-based constraint for the flux estimate.  
		
\begin{figure}[htb]
			\centering
			\begin{adjustwidth}{-0.75in}{-0.5in}
				\begin{subfigure}{0.41\textwidth}
					\includegraphics[width=\textwidth]{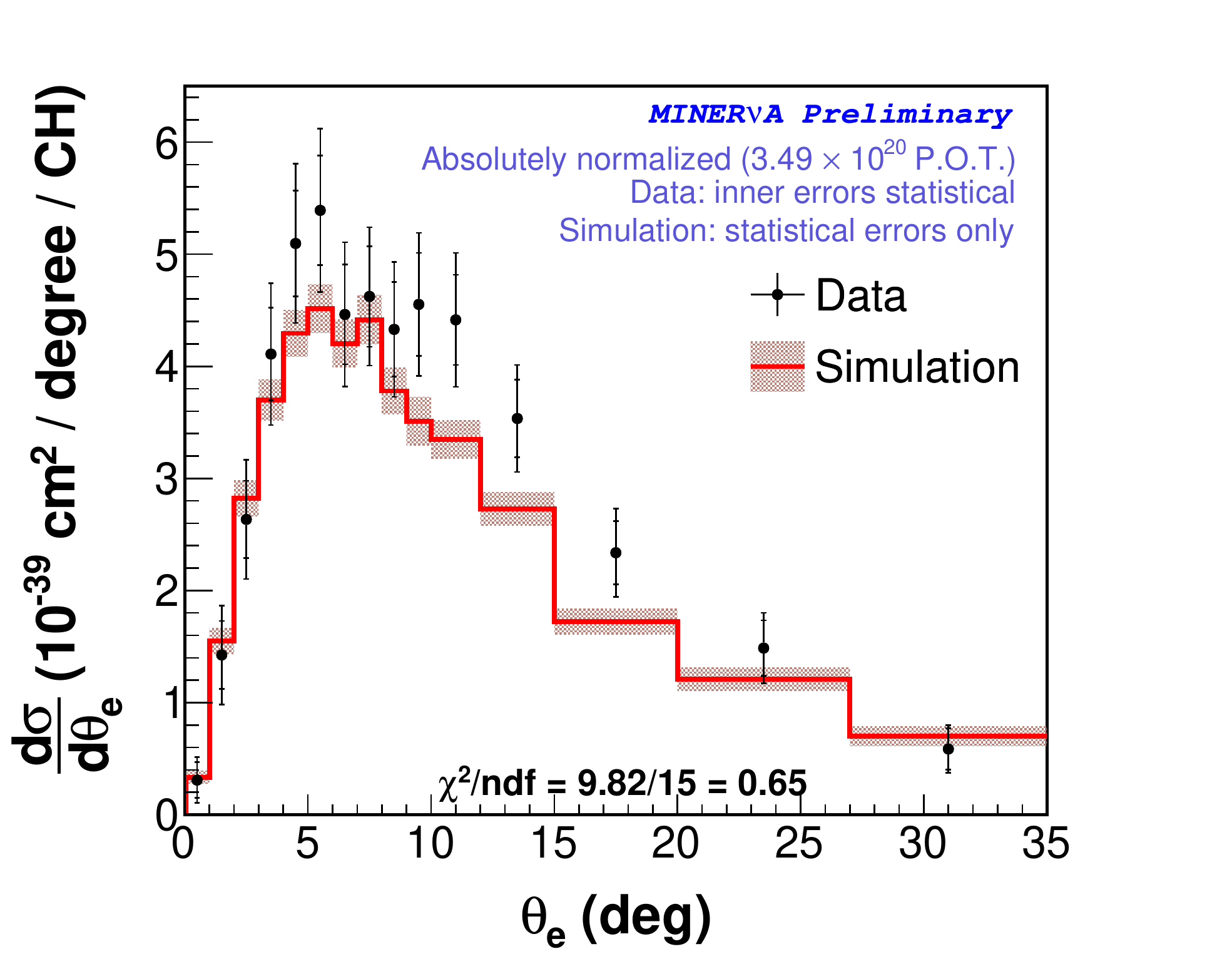}
					\caption{$\theta_{e}$}
				\end{subfigure}%
				\begin{subfigure}{0.41\textwidth}
					\includegraphics[width=\textwidth]{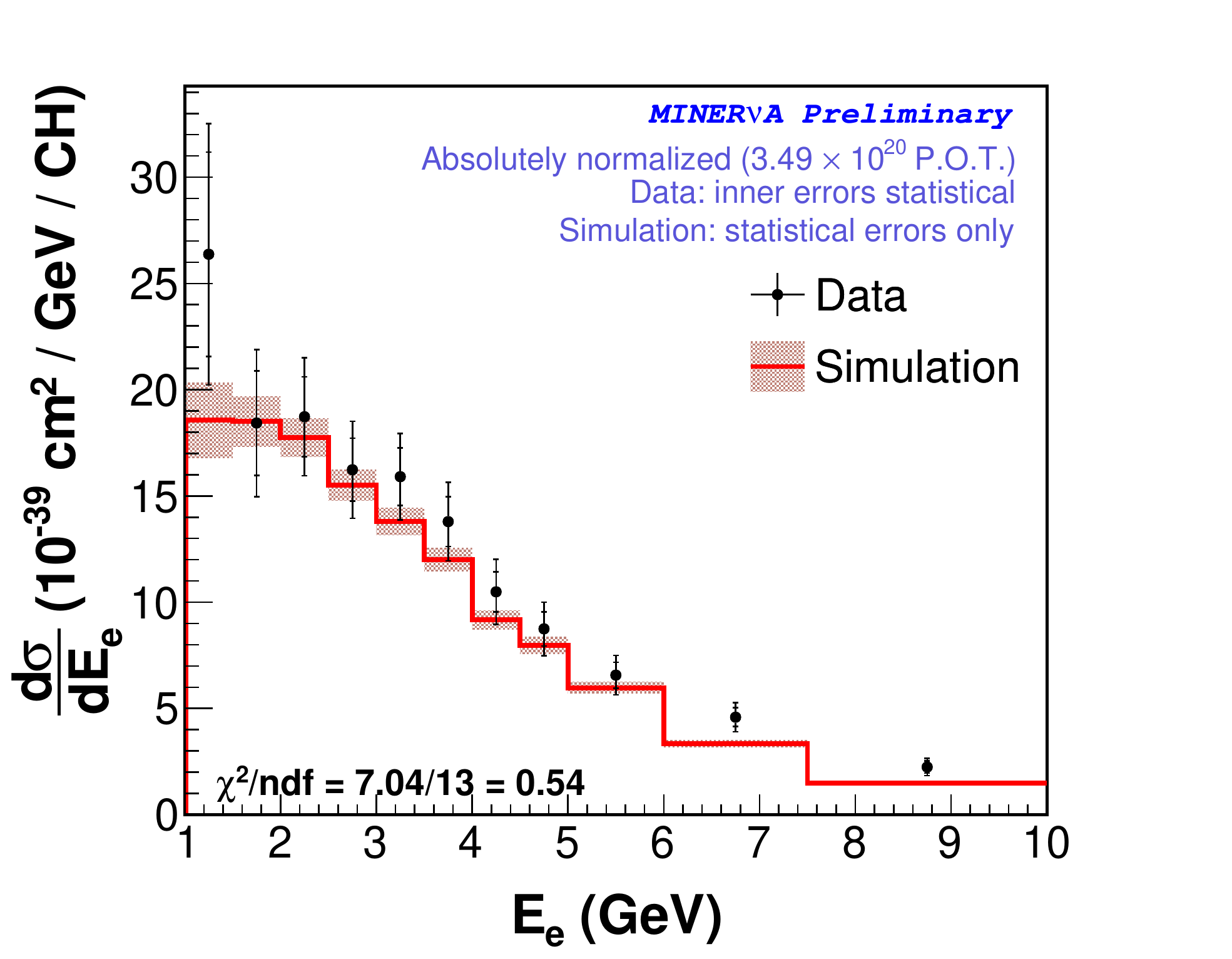}
					\caption{$E_{e}$}
				\end{subfigure}%
				\begin{subfigure}{0.41\textwidth}
					\includegraphics[width=\textwidth]{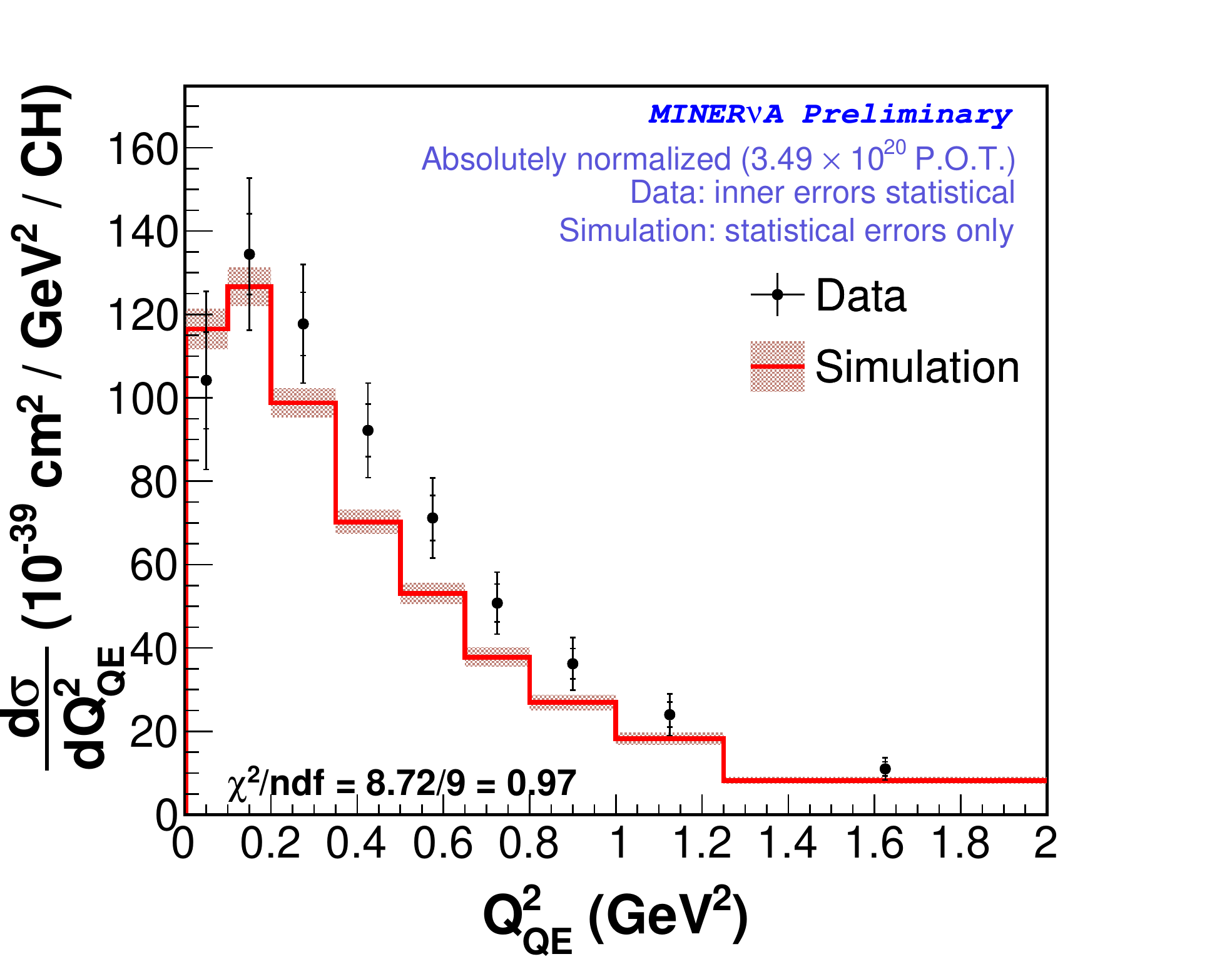}
					\caption{$Q^{2}_{QE}$}
					\label{fig:XS q2}
				\end{subfigure}
			\end{adjustwidth}
			\caption{Differential cross sections.  Inner errors are statistical; outer are statistical added in quadrature with systematic.}
			\label{fig:XSs}
		\end{figure}

\begin{figure}[htb]
			\centering
			\includegraphics[width=0.75\textwidth]{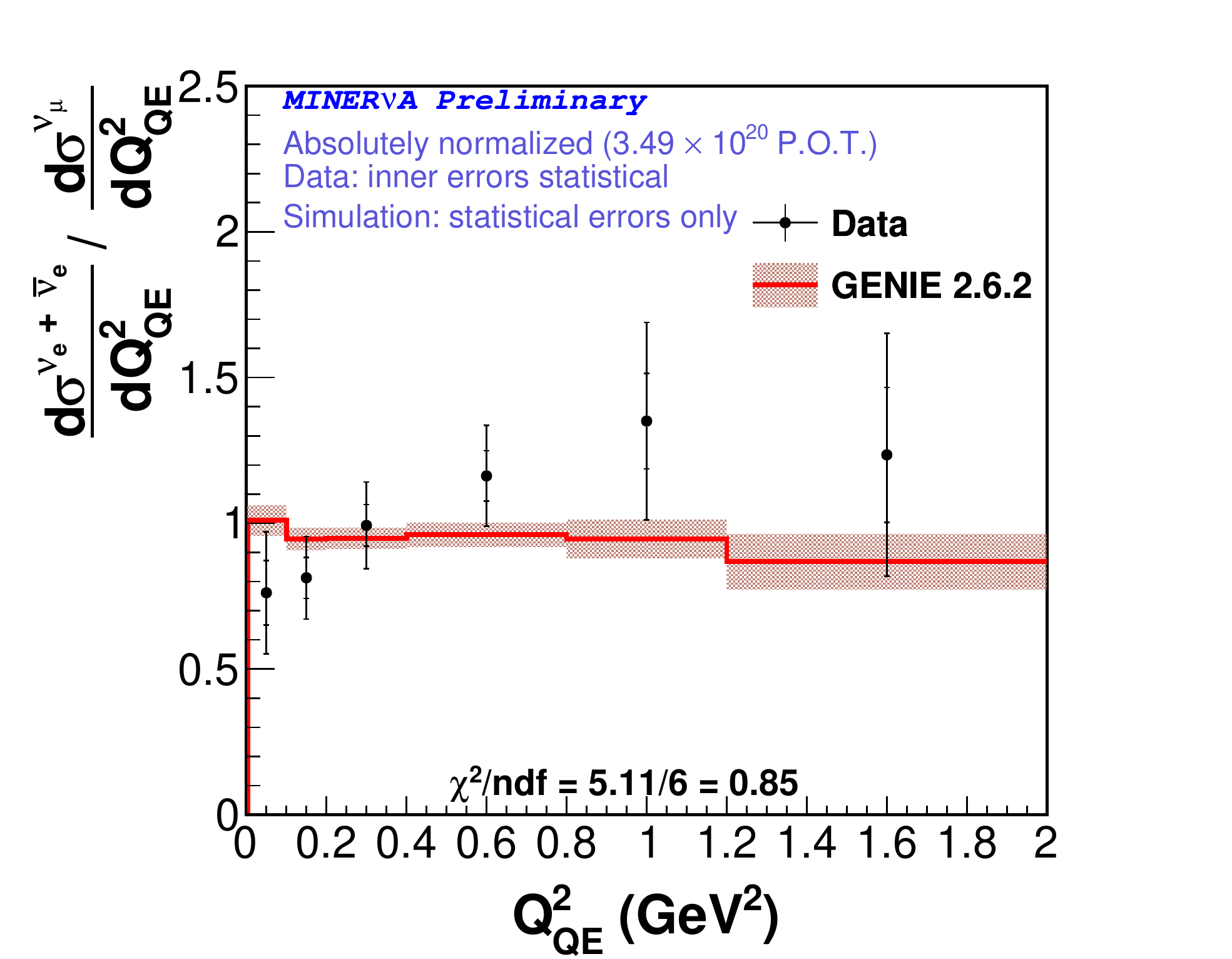}
			\caption{Ratio of $\frac{d\sigma}{dQ^{2}_{QE}}$ for $\nu_{e}$ to that for $\nu_{\mu}$.  Inner errors are statistical; outer are statistical added in quadrature with systematic.}
			\label{fig:ratio}
		\end{figure}

		The cross sections obtained from this procedure are given in fig. \ref{fig:XSs}.  To help understand whether any differences between the model and our data stem from deficiencies in the underlying cross section model itself (which is tuned to $\nu_{\mu}$ scattering data, as noted in the introduction) or differences between $\nu_{e}$ and $\nu_{\mu}$ interactions, we also computed the ratio of the cross section in fig. \ref{fig:XS q2} to a recent \mnv{} measurement of the same cross section for muon neutrinos, which is shown in fig. \ref{fig:ratio}.  We note that $Q^2$-dependent correlated errors, such as that in the electromagnetic energy scale, can cause trends in the data similar to the difference between the prediction and observed shape in $Q^{2}$ in fig. \ref{fig:XS q2} and the apparent upward slope in fig. \ref{fig:ratio}.  When these correlated errors are taken into account, in all cases the data is consistent with the GENIE prediction within $1\sigma$.
		
	\section{Conclusions}
		Though $\nu_{e}$ cross section data is vitally important for neutrino oscillation searches, experimental challenges have prevented extensive measurement of this quantity until recently.  In this first-ever measurement of $\nu_{e}$ CCQE scattering, we find that the electron neutrino cross section predictions of the GENIE generator, based on cross section models tuned to muon neutrino scattering data, are consistent with our measured values within our uncertainties.  This implies that the generator models in their current form are suitable for use by current neutrino oscillation experiments.  However, future experiments, which depend on significantly reducing the influence of cross section systematic uncertainties on their results, may require further data to resolve whether the apparent (but not significant) trends in our result correspond to real discrepancies between the models and nature.

\end{document}